\documentclass[aps,prd,two column]{revtex4-2}

\usepackage{amsmath}
\usepackage{ulem}
\usepackage{graphicx}
\usepackage{epsfig}
\usepackage{bm}
\usepackage{hyperref}
\usepackage{natbib}
\usepackage{caption}
\usepackage{subcaption}
\usepackage{xcolor}

\def\be {\begin{equation}}
\def\ee {\end{equation}}
\def\nn {\nonumber}
\def\bea {\begin{eqnarray}}
\def\eea {\end{eqnarray}}

\begin{document}

\title{Soret and Dufour effects in hot and dense QCD matter}

\author{Kamaljeet Singh}
\author{Kangkan Goswami}
\author{Raghunath Sahoo}
\email{Corresponding Author: Raghunath.Sahoo@cern.ch}
\affiliation{Department of Physics, Indian Institute of Technology Indore, Simrol, Indore 453552, India}

\date{\today}
\begin{abstract}
The gradients act as invisible engines of transport, converting microscopic imbalances into macroscopic flows, and thus providing deep insights into the dynamics of physical systems. Thermal gradients do not merely drive the flow of heat, but they also set the microscopic constituents of the system into motion. In such scenarios, the constituents of the system not only transport energy but also diffuse collectively under the influence of these gradients. 
For the very first time, we present a first-principles investigation of the Soret and Dufour effects in hot and dense quantum chromodynamics (QCD) matter. We use the relativistic Boltzmann transport equation under the relaxation time approximation.  By incorporating baryon chemical potential and temperature gradients into the kinetic theory framework, we derive explicit expressions for the Dufour coefficient, which quantifies the heat flow due to concentration gradients, and the Soret coefficient, which describes the particle diffusion induced by thermal gradients. These coupled-transport phenomena are traditionally studied in multi-component classical systems at low energy scales. In this study, we follow quasiparticle models for the deconfined phase and the hadron resonance gas model for the confined hadronic phase in the context of heavy-ion collisions. 
This study provides novel insights into the thermo-diffusion and diffusion-thermo phenomena and opens avenues for incorporating such effects in hydrodynamic modeling and transport simulations of QCD matter. 
\end{abstract}

\maketitle

\section{Introduction}
Transport phenomena after the Big Bang played a key role in shaping the early universe, helping it cool from a quark-gluon plasma (QGP) phase to a hadron resonance gas (HRG) phase and setting the stage for the formation of cosmic structures. High-energy heavy-ion collisions recreate, on a miniature scale, the extreme temperature and density conditions that existed microseconds after the Big Bang, enabling the study of the QGP medium and its transition to hot and dense hadronic matter~\cite{Busza:2018rrf}. Understanding the transport properties of the hot and dense strongly interacting quantum chromodynamics (QCD) matter created in ultra-relativistic heavy-ion collisions stands as one of the most intriguing and captivating phenomena for the scientific community, offering deep insights into its intricate dynamics. The experimental facilities at the Relativistic Heavy Ion Collider (RHIC) and the Large Hadron Collider (LHC) provide a unique platform to study the dynamics of strongly interacting matter created by colliding heavy nuclei at ultrarelativistic speeds. In these collisions, the early-formed deconfined phase expands and cools with time and is finally confined into hadrons, which further undergo the freezeout stages prior to the free streaming of particles towards the detectors~\cite{Brewer:2019oha,Lysenko:2024hqp}. The final state particles play a key role in understanding the underlying physics of the evolution of the fireball~\cite{Sahoo:2021aoy}. Beyond experimental observations, extensive phenomenological studies offer detailed descriptions of the microscopic dynamics and transport phenomena governing QCD matter. In the last decade, the study of thermal and electrical conductivity of QCD matter within different models has provided a good phenomenological understanding of transport phenomena~\cite {Gupta:2003zh, Ding:2010ga, Cassing:2013iz, FernandezFraile:2005bew, Marty:2013ita, Singh:2023pwf, Dey:2020awu, Ghosh:2019ubc, Nam:2012sg, Hattori:2016cnt, Hattori:2016lqx, Harutyunyan:2016rxm, Kerbikov:2014ofa, Feng:2017tsh, Wang:2020qpx, Tuchin:2012mf, Singh:2023ues}. Furthermore, various realistic effects, such as transient electromagnetic fields and vorticity-induced rotation in the medium, have been studied~\cite{Buividovich:2010tn, Nam:2012sg, Hirono:2012rt, Singh:2023pwf, Singh:2023ues, Dwibedi:2025boz, Padhan:2024edf}. Also, the studies have extended from transport of electric currents to include baryon, strangeness, and off-diagonal strange–baryon currents~\cite{Greif:2017byw, Fotakis:2019nbq, Fotakis:2021diq, Das:2021bkz}. With the recent experimental developments, in the charm sector, the charm current and its off-diagonal components have been studied in the hadronic medium as well~\cite{Goswami:2024hfg}. In the context of heavy-ion collisions, the authors in Ref.~\cite{Singh:2024emy, Singh:2025rwc, Singh:2025geq, Singh:2025oja} studied the leading and higher-order thermoelectric and magneto-thermoelectric transport coefficients of the QCD matter. These leading-order coefficients include the Seebeck, magneto-Seebeck, and Nernst effects, while the higher-order contributions involve the Thomson, magneto-Thomson, and transverse Thomson effects, offering a more comprehensive understanding of electric charge and energy transport in the strongly interacting medium. The study of the thermoelectric figure of merit across the deconfinement phase transition is studied in Ref.~\cite{Singh:2025zkr}. Unlike the conventional Seebeck effect~\cite{Singh:2024emy,Li:2019bgc,PhysRevB.105.235116, Abhishek:2020wjm}, where a temperature gradient induces only an electric current in the QCD medium, such gradients can also drive a net baryon flow, leading to baryon diffusion. This unique coupling between electric and baryon charge currents under the temperature gradients provides valuable insights into the coupled-transport properties of strongly interacting matter and plays a key role in understanding the evolution of the baryon-rich medium created in heavy-ion collisions.

 The coupled-transport phenomena offer a unique window into the microscopic dynamics that govern the macroscopic behavior of the system. Such interesting coupled phenomena are the Dufour and Soret effects~\cite{awad2010dufour,narayana2008soret,bilal2016three,mahdy2010soret,shaheen2021soret}. The Dufour effect refers to the generation of a heat current due to a concentration gradient, also known as the diffusion-thermo effect. The Soret effect describes a phenomenon where a temperature gradient leads to the diffusion of the constituents of the medium, hence called as thermo-diffusion effect. These effects have numerous practical applications across various fields, including chemical and geophysical engineering, which have attracted significant attention from researchers over the years~\cite{rasool2020consequences}. For instance, Ref.~\cite{jiang2020physical} reported notable findings on the simultaneous transport of heat and mass through species inter-diffusion processes under the direct influence of the Soret and Dufour effects, particularly in natural convection driven by evaporation and condensation phenomena. Similarly, Ref.~\cite{liu2019multiple} incorporated multiple-relaxation mechanisms within the lattice Boltzmann framework to investigate dual-diffusive natural convective flows, demonstrating that the combined influence of Soret and Dufour effects plays a critical role in governing these transport processes. In the Ref.~\cite{garcia2007dufour}, analytical expressions for the Soret and Dufour effects were derived, highlighting their significant contribution to coupled-transport phenomena in magnetized electrodynamics plasmas. Studying these phenomena in the domain of high-energy physics opens an exciting opportunity for understanding the coupled-transport dynamics of QCD matter. 
In the context of relativistic hydrodynamics, a thermo-diffusion contribution to the baryon diffusion current emerges naturally once the baryon sector is formulated within a causal second-order (Israel–Stewart) framework. In the first-order Navier–Stokes theory used, for example, in Ref.~\cite{Kovtun:2012rj}, the baryon current is driven by only a single thermodynamic force, which prevents the appearance of an independent response to a temperature gradient. Any temperature dependence is implicitly folded into a single transport coefficient. By contrast, second-order hydrodynamics, as mentioned in Ref.~\cite{Du:2021zqz}, provides a relaxation equation in which gradients of temperature and baryon density act as distinct thermodynamic forces with their own transport coefficients. This causal structure naturally allows for a thermo-diffusion (Soret-type) term that is dynamically generated rather than fixed by first-order constitutive relations. Moreover, the Ref.~\cite{Monnai:2012jc}, explicitly identifies a baryon--heat cross conductivity that generates the coefficient of thermo-diffusion in a second-order framework. That study demonstrates that a temperature gradient can independently drive baryon diffusion when off-diagonal transport coefficients are included, in full agreement with Onsager reciprocity. 
 
 In the realm of ultra-relativistic heavy-ion collisions, the local temperature gradient naturally arises as the system expands and cools, while concentration gradients appear due to the system being slightly away from equilibrium. Incorporating Dufour and Soret effects into the study of QCD matter would provide a new dynamical phenomenon that would affect the evolution of the medium.  The Dufour effect influences the heat current in the medium due to the gradient in the baryon chemical potential, which can essentially affect the cooling of the medium. Moreover, the Soret coefficient modifies the diffusion current due to the existing temperature gradient and can eventually affect the diffusion of the baryons in the medium, which can introduce an effect on experimental observables such as net proton number fluctuations and charge particle spectra. Thus, a study of these coupled-transport effects in the domain of heavy-ion collisions is important and could provide valuable insights into the interplay between heat and particle transport in QCD matter. 


\section{Framework for the QCD Equation of State}\label{sec:formalism}

This section outlines the phenomenological models used to describe the QCD matter created in heavy-ion collisions. We briefly present the quasiparticle model, QPM, for the QGP phase and the hadron resonance gas HRG model for the hadronic phase to characterise the equation of state (EoS). 

\subsection{Quasiparticle Model}
\label{model1}

To model the quark-gluon plasma phase, we follow a quasiparticle model approach originally introduced by Gorenstein and Yang~\cite{Gorenstein:1995vm}. In this framework, the medium is described as a gas of quasi-free partons possessing medium-dependent effective masses that encapsulate interaction effects. The temperature-dependent effective mass \( m(T) \) accounts for the non-perturbative dynamics of the QGP medium, while a temperature-dependent bag constant ensures thermodynamic consistency by reflecting vacuum contributions.

The energy of $i$th parton with momentum \( k_i \) is given by the dispersion relation
\begin{equation}
    \omega_i^2(k_i, T) = \vec{k}_i^2 + m_i^2(T),
\end{equation}
where \( \omega_i(T) \) is the energy, and  \( m_i(T) \) is the total effective mass of the $i$th parton. For the $i$th quark, the effective mass is parametrized as
\begin{equation}
    m_i^2 = m_{i0}^2 + \sqrt{2}m_{i0}m_{iT} + m_{iT}^2,
\end{equation}
with \( m_{i0} \) representing the bare quark mass and \( m_{iT} \) the thermal mass contribution, given by
\begin{equation}
    m_{iT}^2(T,\mu_B) = \frac{N_c^2 - 1}{8N_c}\left( T^2 + \frac{\mu_B^2}{9\pi^2} \right)g^2(T).
\end{equation}

The gluon effective mass is described by
\begin{equation}
    m_g^2(T,\mu_B) = \frac{N_c}{6} g^2(T) T^2 \left(1+\frac{N_f + \mu_B^2/(\pi^2 T^2)}{6}\right),
\end{equation}
where \( N_c \) is the number of color degrees of freedom, \( N_f \) is the number of flavor degrees of freedom and \( g^2(T) = 4\pi \alpha_s(T) \), with \( \alpha_s(T) \) denoting the running strong coupling constant.

The relaxation time \( \tau_R \) of quarks is taken to be momentum-independent and is given by~\cite{Hosoya:1983xm}
\begin{equation} \label{tauqgp}
    \tau_R = \frac{1}{5.1\,T\,\alpha_s^{2}\log(1/\alpha_s)[1 + 0.12(2N_f + 1)]},
\end{equation}
where a fixed value \( \alpha_s = 0.5 \) is used for numerical evaluation.

\subsection{Ideal Hadron Resonance Gas Model}
\label{model2}

To describe the hadronic phase of QCD matter, particularly at temperatures below the crossover transition, we use the Ideal Hadron Resonance Gas model~\cite{Singh:2025oja}. This statistical model considers a non-interacting gas composed of all known hadrons and resonances listed in the Particle Data Group (PDG)~\cite{ParticleDataGroup:2008zun}. Despite the non-interacting considerations, the inclusion of resonances effectively mimics the strong interactions among hadrons based on the idea that interactions in the medium can be approximated by
resonance formation.

The grand canonical partition function for the $i$th hadron species is given by~\cite{Pradhan:2022gbm}
\begin{equation}
\label{eq:partition}
\ln Z^{id}_i = \pm \frac{Vg_i}{2\pi^2} \int_{0}^{\infty} \vec{k}_i^2 d\lvert{\vec{k_i}\rvert} \ln\left[1\pm \exp\left(-\frac{\omega_i - b_i\mu_B}{T}\right)\right],
\end{equation}
where \( g_i \) is the degeneracy factor, \( \omega_i = \sqrt{\vec{k}_i^2 + m_i^2} \) is the single-particle energy, \( b_i \) is the baryon number, and \( \mu_B \) is the baryon chemical potential. The upper and lower signs correspond to fermions (baryons) and bosons (mesons), respectively.

From the partition function, the following thermodynamic quantities can be derived
\begin{align}
P^{id}_i(T,\mu_i) &= \pm \frac{Tg_i}{2\pi^2} \int_{0}^{\infty} \vec{k}_i^2 d\lvert{\vec{k_i}\rvert} \ln\left[1\pm \exp\left(-\frac{\omega_i - b_i\mu_B}{T}\right)\right], \\
\varepsilon^{id}_i(T,\mu_i) &= \frac{g_i}{2\pi^2} \int_{0}^{\infty} \frac{\omega_i\, \vec{k}_i^2 d\lvert{\vec{k_i}\rvert}}{\exp\left(\frac{\omega_i - b_i\mu_B}{T}\right)\pm1},\\
n^{id}_i(T,\mu_i) &= \frac{g_i}{2\pi^2} \int_{0}^{\infty} \frac{\vec{k}_i^2 d\lvert{\vec{k_i}\rvert}}{\exp\left(\frac{\omega_i - b_i\mu_B}{T}\right)\pm1}.
\end{align}

The thermal relaxation time \( \tau_R^i \) for hadrons is computed by thermally averaging the energy-dependent relaxation time over the equilibrium distribution. It is expressed as~\cite{Das:2020beh}
\begin{equation}\label{tauhrg}
    (\tau_R^i)^{-1} = \sum_j n_j \langle\sigma_{ij}v_{ij}\rangle,
\end{equation}
where the thermal average of the cross-section times relative velocity between $i$th and $j$th hadrons is given by
\begin{align}
\langle\sigma_{ij}v_{ij}\rangle &= \frac{\sigma}{8T\,m_i^2 m_j^2\,\mathcal{K}_2(m_i/T)\,\mathcal{K}_2(m_j/T)} \int_{(m_i + m_j)^2}^{\infty} ds \nonumber \\
&\quad \times \frac{[s - (m_i - m_j)^2]}{\sqrt{s}} [s - (m_i + m_j)^2] \mathcal{K}_1(\sqrt{s}/T),
\end{align}
where \( \sigma = \pi (r_i+r_j)^2 \) is the hard-sphere scattering cross-section (assumed independent of \( T \) and \( \mu_B \)), $\sqrt{s}$ is the center of mass energy, and \( \mathcal{K}_1 \), \( \mathcal{K}_2 \) are modified Bessel functions of the second kind. In our current study, we consider the same radius for all mesons as 0.2 fm, and for all baryons, it is 0.62 fm

\section{coupled-transport coefficients in QCD Matter}
\label{formalism2}
In a medium with a non-zero net baryon number, the transport of baryons and energy (heat) can be described by linear response theory. The diffusion current \( \vec{j} \) and the heat current \( \vec{I} \) are expressed using Fick's law generalised for coupled-transport phenomena as
\begin{align}
\begin{pmatrix}\label{matrix-1}
\vec{j} \\
\vec{I}
\end{pmatrix}
=
- \begin{pmatrix}
D & S_T \\
D_F & \kappa
\end{pmatrix}
\begin{pmatrix}
\vec{\nabla} \mu_B \\
\vec{\nabla} T
\end{pmatrix}
\end{align}
where \( D \) is the baryon diffusion coefficient, \( S_T \) is the Soret coefficient, \( \kappa \) is the coefficient of thermal conductivity, and \( D_F \) is the Dufour coefficient. These expressions capture the interplay between chemical potential and thermal gradients in driving both baryons and heat transport in the medium.

To evaluate the coupled-transport coefficients in QCD matter, we follow the kinetic theory framework. We begin by decomposing the single-particle distribution function for species \( i \) into its equilibrium part and a small deviation
\[
f_i = f_i^0 + \delta f_i,
\]
where \( f_i^0 \) represents the equilibrium distribution and \( \delta f_i \) accounts for slight deviation from equilibrium. At equilibrium, the distribution for the $i$th species is expressed as
\begin{equation}\label{Dis-f-new}
f_i^0 = \frac{1}{\exp\left[\frac{\omega_i - b_i \mu_B}{T}\right] \pm 1}~,
\end{equation}
with \(\omega_i\) being the energy of the particle, \(\mu_B\) the baryon chemical potential, and \(b_i\) the baryon number ({\it e.g.}, \(b_i=1\) for baryons, \(b_i=-1\) for antibaryons, and \(b_i=0\) for mesons). The plus and minus signs correspond to fermions and bosons, respectively.

In the absence of any external force, in the local rest frame, the Boltzmann transport equation (BTE) under the relaxation time approximation (RTA) for species \(i\) can be written as~\cite{Das:2020beh, Singh:2023pwf}
\begin{equation}\label{BTE-new}
\vec{v}_i \cdot \vec{\nabla} f_i 
= -\frac{\delta f_i(\vec{x}_i, \vec{k}_i)}{\tau_R^i}~.
\end{equation}
Here, \(\tau_R^i\) denotes the relaxation time for species \(i\), and \(\vec{v}_i = \vec{k}_i/\omega_i\) is the particle velocity. Additionally, the spatial gradient of \( f_i^0 \) can be recast using
\begin{equation}\label{f0-grad-Gibbs-new}
\vec{\nabla} f_i^0 = -\frac{\partial f_i^0}{\partial \omega_i} \Bigg\{\frac{\left( \omega_i - b_i \, \mu_B \right)}{T} \vec{\nabla}T + b_i \vec{\nabla}\mu_{B} \Bigg\}~,
\end{equation}
At leading order, an ansatz for the nonequilibrium part can be chosen as~\cite{ Singh:2023pwf, Singh:2025zkr,Das:2019pqd}:
\begin{equation}\label{delta-f-new}
\delta f_i = (\vec{k}_i \cdot \vec{\Omega}) \frac{\partial f_i^0}{\partial \omega_i}~,
\end{equation}
where the vector \(\vec{\Omega}\) encapsulates the influence of the driving forces. It is natural to express \(\vec{\Omega}\) as a linear combination of the perturbed fields
\begin{equation}\label{Omega-new}
\vec{\Omega} = \alpha_1\, \vec{\nabla}T + \alpha_2\, \vec{\nabla}\mu_B~,
\end{equation}
with the unknown coefficients \(\alpha_1\) and \(\alpha_2\) that quantify the strength of the electric field and temperature gradient in pushing the system out of equilibrium.

Substituting Eq.~\eqref{Omega-new} into \eqref{delta-f-new}, one obtains the following expression for the deviation \(\delta f_i\)
\begin{equation}\label{delta-f-final-new}
\delta f_i = \omega_i  \left[ \, \alpha_1(\vec{v}_i\cdot\vec{\nabla}T) + \alpha_2(\vec{v}_i\cdot\vec{\nabla}\mu_B) \right]\frac{\partial f_i^0}{\partial \omega_i}~.
\end{equation}

Hence, the relativistic Boltzmann equation in Eq.~\eqref{BTE-new} can be expressed as
\begin{align}
\frac{\left( \omega_i - b_i \, \mu_B \right)}{T} (\vec{v}_i\cdot\vec{\nabla}T) + b_i (\vec{v}_i\cdot\vec{\nabla}\mu_B) = \frac{\omega_i}{\tau_R^i}  [ \, \alpha_1(\vec{v}_i\cdot\vec{\nabla}T)\nn\\
+ \alpha_2(\vec{v}_i\cdot\vec{\nabla}\mu_B)]
\end{align}
After comparing the coefficients on both the sides, we get
\begin{align}
\alpha_1 = \frac{\tau_R^i\left( \omega_i - b_i \, \mu_B \right)}{T\omega_i},~~~~~~
\alpha_2  = \frac{b_i\tau_R^i}{\omega_i}
\end{align}
Hence,
\begin{equation}\label{delta}
\delta f_i = \frac{\vec{k}_i}{\omega_i}\tau_R^i\left[ \, \frac{\left( \omega_i - b_i \, \mu_B \right)}{T}\vec{\nabla}T + b_i\vec{\nabla}\mu_B \right] \frac{\partial f_i^0}{\partial \omega_i} ~.
\end{equation}
Following the kinetic theory, one can write the particle diffusion current and heat current as
\begin{align}
\vec{j} &= \sum_i g_i\int \frac{d^3\lvert{\vec{k_i}\rvert}\vec{k}_i}{(2\pi)^3 \omega_i}   \delta f_i, \nn\\
\vec{I} &= \sum_i g_i \int \frac{d^3\lvert{\vec{k_i}\rvert}~\vec{k}_i}{(2\pi)^3 \omega_i} (\omega_i - b_i\mu_B)  \, \delta f_i.
\end{align}
Here, $g_i$ is the degeneracy for the $i$th species of the medium. After substituting the $\delta f_i$ from Eq.~\eqref{delta} in the above equation and comparing with the equations defined in the matrix.~\eqref{matrix-1}, we get
\begin{itemize}

\item {Diffusion coefficient (baryon diffusion due to baryon chemical potential gradient):}
\begin{align}\label{Eq:Diffusion}
D = \sum_i g_i ~\frac{1}{3T} \int \frac{d^3\lvert{\vec{k_i}\rvert}}{(2\pi)^3} 
\frac{\vec{k}_i^2}{\omega_i^2}~b_i \tau_R^i f_i^{0} (1 \pm f_i^{0}).
\end{align}

\item {Soret coefficient (baryon diffusion due to temperature gradient):}
\begin{align}\label{Eq:Soret}
S_{T} = \sum_i g_i~ \frac{1}{3T^2} \int \frac{d^3\lvert{\vec{k_i}\rvert}}{(2\pi)^3} 
\frac{\vec{k}_i^2}{\omega_i^2} (\omega_i - b_i\mu_B) \tau_R^i f_i^{0} (1 \pm f_i^{0}).
\end{align}

\item {Dufour coefficient (heat flow due to baryon chemical potential gradient):}
\begin{align}\label{Eq:Dofour}
D_{F} =\sum_i g_i~ \frac{1}{3T} \int \frac{d^3\lvert{\vec{k_i}\rvert}}{(2\pi)^3} 
\frac{\vec{k}_i^2}{\omega_i^2} (\omega_i - b_i\mu_B)~b_i\tau_R^i f_i^{0} (1 \pm f_i^{0}).
\end{align}

\item {Thermal conductivity (heat flow due to temperature gradient):}
\begin{align}\label{Eq:kappa}
\kappa = \sum_i g_i~ \frac{1}{3T^2} \int \frac{d^3\lvert{\vec{k_i}\rvert}}{(2\pi)^3} 
\frac{\vec{k}_i^2}{\omega_i^2} (\omega_i - b_i\mu_B)^2 \tau_R^i f_i^{0} (1 \pm f_i^{0}).
\end{align}

\end{itemize}

These transport coefficients collectively describe the coupled flow of heat and baryon charge in the QCD medium. These coefficients incorporate the statistical weights, baryon numbers, effective energies, relaxation times, and equilibrium distributions, providing a comprehensive description of transport phenomena in the strongly interacting medium.

\begin{figure*}
	\centering
    {\includegraphics[scale=0.40]{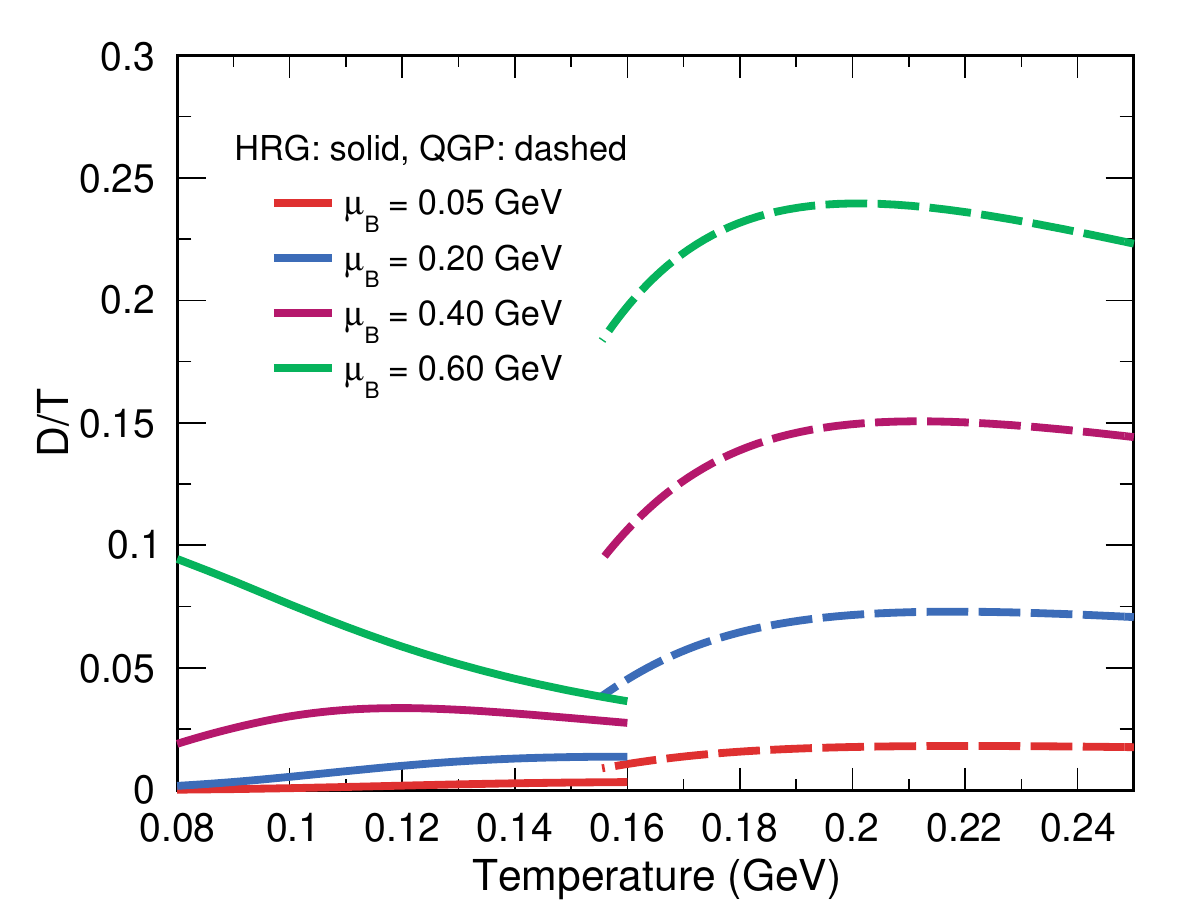}}
    \includegraphics[scale=0.40]{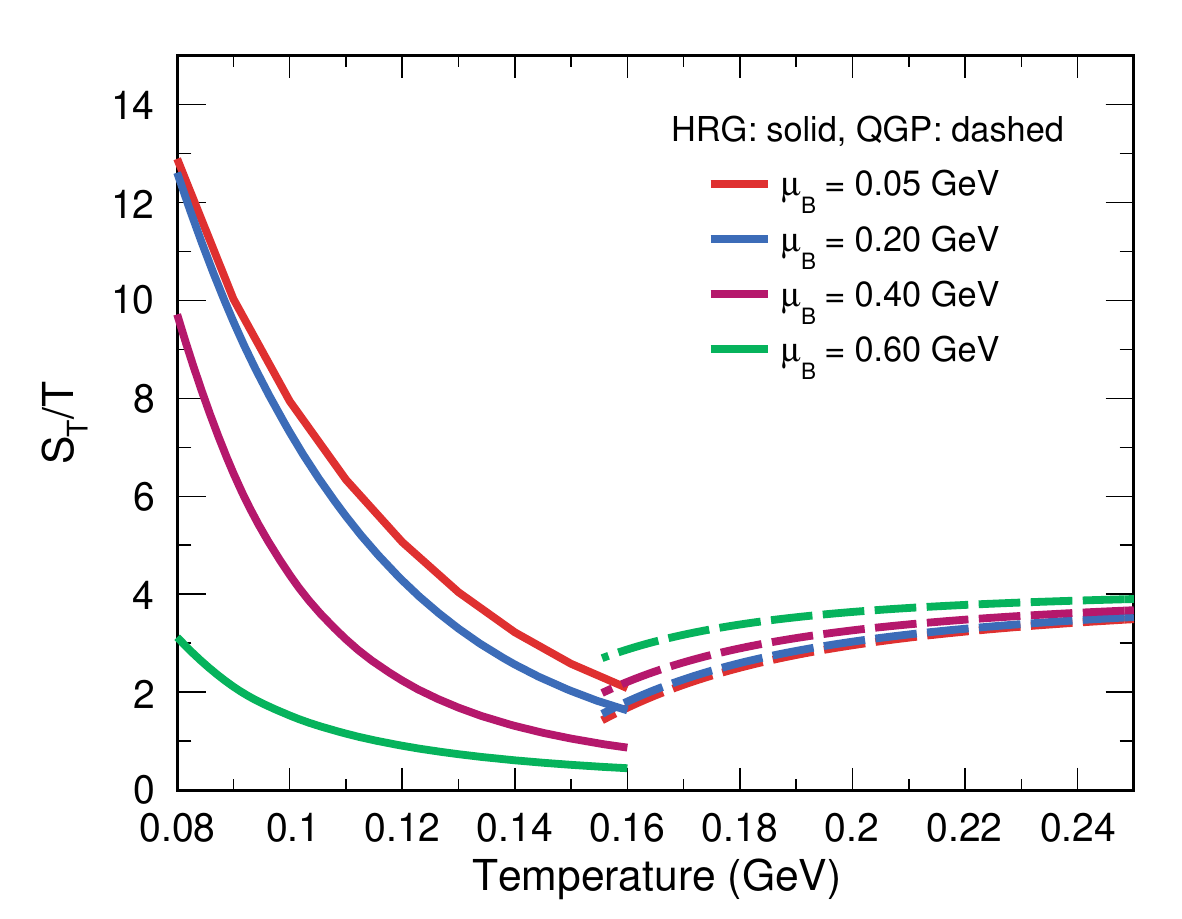}    
    \includegraphics[scale=0.40]{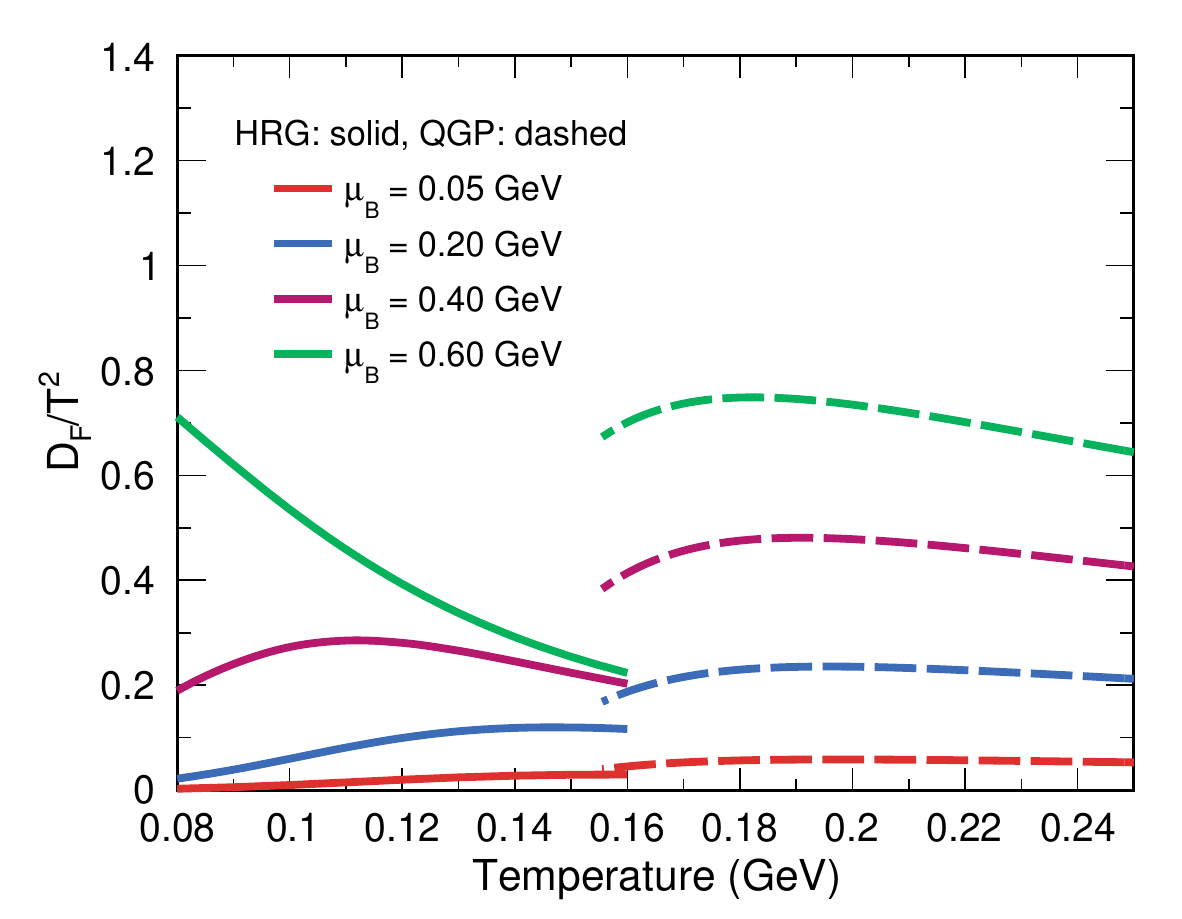}
    {\includegraphics[scale=0.40]{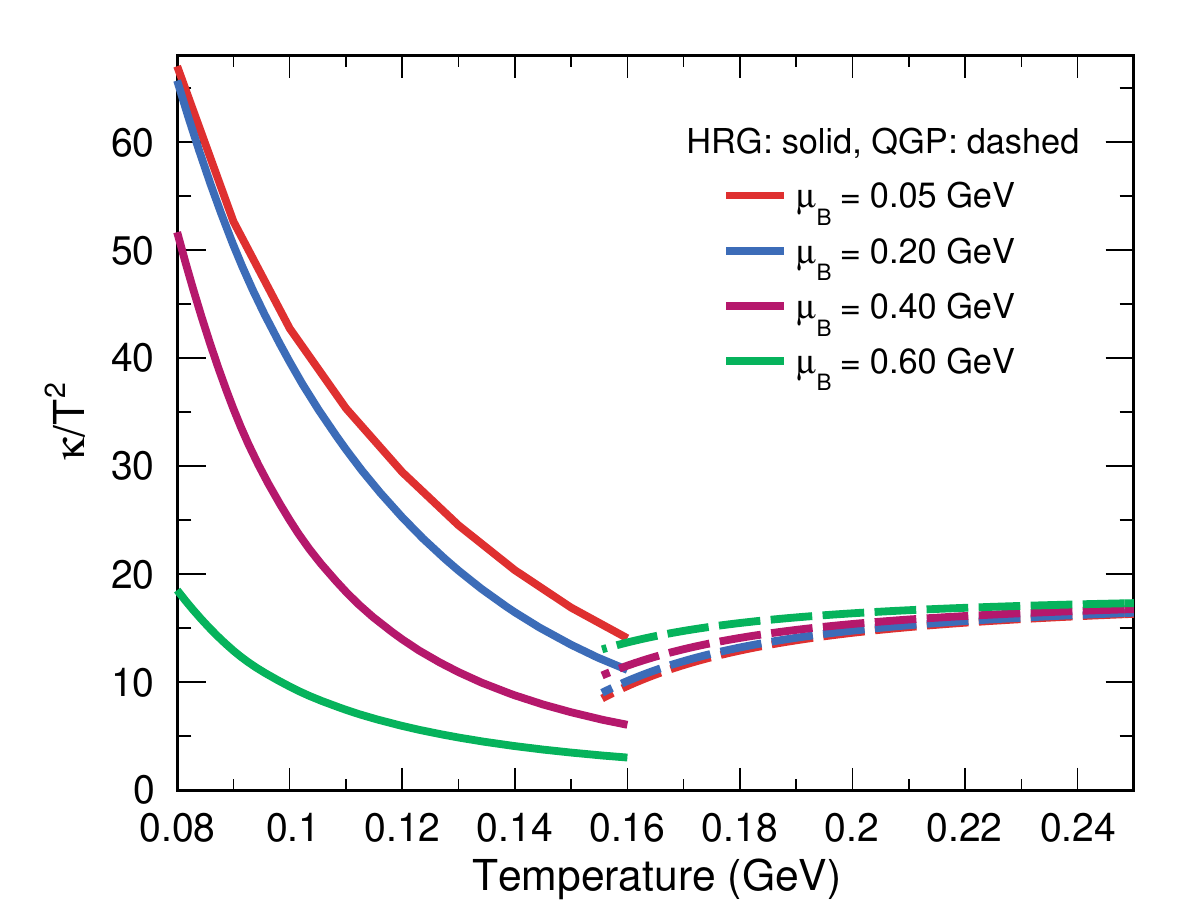}}
	\caption{
Temperature dependence of scaled coefficients of coupled-transport in the hadron resonance gas (HRG) and quark-gluon plasma (QGP) phases at different values of baryon chemical potential $\mu_{B}$. Shown are scaled (top left) diffusion coefficient ($D/T$), (top right) Soret coefficient ($S_{T}/T$), (bottom left) Dufour coefficient ($D_{F}/T^{2}$), and (bottom right) thermal conductivity ($\kappa/T^{2}$). HRG results are plotted with solid lines and QGP results with dashed lines. Each colour corresponds to a fixed $\mu_{B}$ value.
}\label{fig:thermo-diffusion_coeffs}
\end{figure*}
\section{Results and Discussion}\label{Sec-results}
In this study, we have calculated the coupled-transport coefficients for the QCD matter using the QPM model for the deconfined phase and the HRG model for the confined phase. Figure~\ref{fig:thermo-diffusion_coeffs} shows the temperature dependence of the scaled coupled-transport coefficients in the hadron resonance gas (solid lines) and quark-gluon Plasma (dashed lines) phases for different values of the baryon chemical potential, $\mu_B = 0.05,\, 0.20,\, 0.40$, and $0.60~\mathrm{GeV}$. The four panels correspond to: 
(top left) the scaled diffusion coefficient ($D/T$), 
(top right) the scaled Soret coefficient ($S_T/T$), 
(bottom left) the scaled Dufour coefficient ($D_F/T^2$),
and (bottom right) the scaled thermal conductivity ($\kappa/T^2$).
The mathematical forms of these coefficients are mentioned in Eqs.~\eqref{Eq:Diffusion}-\eqref{Eq:kappa}. The integrands of these integrals contain several key terms with distinct physical meanings. The particle momentum squared, $\vec{k}_i^2$, scaled by the squared particle energy, $\omega_i^2$, represents the kinetic contribution of the particles to the transport processes. The relaxation time, $\tau_R^i$, characterizes the interaction strength or mean free time between successive collisions for the $i$th species, dictating how effectively each particle species participates in the transport phenomena. The equilibrium occupation factor, $f_i^0 (1 \pm f_i^0)$, accounts for the quantum statistical effects, where the '$+$' sign corresponds to bosons and the '$-$' sign corresponds to fermions, thereby modifying the overall transport behavior according to the underlying quantum statistics of the medium. The Eqs.~\eqref{tauqgp} and \eqref{tauhrg} represent the relaxation time of the particle species for the QGP and the HRG phases, respectively. For the QGP phase, the $\tau_R^i$ dependence only on the temperature, and it decreases as the temperature increases. Whereas for the HRG phase, the relaxation time $\tau_R$ exhibits a strong dependence on temperature and baryon chemical potential due to its inverse relation with the product of particle density, relative velocity, and interaction cross section. At low temperatures, the densities of most hadrons are exponentially suppressed, resulting in very large relaxation times, indicating infrequent collisions. As the temperature increases toward the crossover region, the hadronic densities grow rapidly, particularly for light mesons like pions, leading to a significant decrease in $\tau_R$. The baryon chemical potential further decreases $\tau_R$ in a baryon-rich medium due to the higher number density. Consequently, in the HRG phase, $\tau_R$ evolves from a dilute, weakly interacting regime at low $T$ to a more collisional, rapidly relaxing medium near the QCD crossover.

The diffusion coefficient in Eq.~\eqref{Eq:Diffusion} quantifies the baryon flux generated by a gradient in the baryon chemical potential $\vec\nabla \mu_B$, indicating how efficiently baryons redistribute in response to $\vec\nabla \mu_B$. It is directly dependent on the factor of $\tau_R^i~f_i^0 (1 \pm f_i^0)$. Hence, this product describes the overall behaviour of the diffusion coefficient. It is to be noted that the Eq.~\eqref{Eq:Diffusion} also carries an explicit factor of baryon number, $b_{i}$, in the product to factor $\tau_R^i~f_i^0 (1 \pm f_i^0)$, hence for bosons, i.e., gluons in QGP phase and mesons in HRG phase, the diffusion coefficient vanishes. In the limit of vanishing baryon chemical potential, the net baryon density becomes zero because of the equal population of baryons and antibaryons carrying opposite baryon numbers, leading to the complete suppression of the diffusion coefficient. Consequently, with increasing $\mu_{B}$, the baryon densities increase through the Boltzmann factor, $\exp(\mu_{B}/T)$, which leads to a significant rise in $D/T$ with $\mu_{B}$, especially in the moderate-temperature region. A larger diffusion coefficient indicates that baryons can diffuse more easily through the medium under a baryon chemical potential gradient. Similarly, the Dufour coefficient in Eq.~(\ref{Eq:Dofour}) includes the same $b_{i}$ dependence as the diffusion coefficient but is further weighted by the factor of $(\omega_{i}-b_{i}\mu_{B})$, which significantly enhances its sensitivity to the baryon number density in the medium. This dependence makes the Dufour effect a predominantly baryon-driven transport phenomenon in the hadronic medium. At very low baryon chemical potential, the net baryon density is strongly suppressed, which results in negligible values of $D_{F}/T^{2}$. However, as $\mu_{B}$ increases, the baryon density grows exponentially, and the Dufour coefficient rises sharply, showing the strongest enhancement among all the transport coefficients considered in this study. This behavior highlights the fact that the Dufour effect, which represents the heat flow driven by chemical potential gradients, becomes increasingly significant in baryon-rich environments, such as the lower-energy heavy-ion collisions. Moreover, the additional energy weighting $(\omega_{i}-b_{i}\mu_{B})$ implies that heavier baryonic resonances contribute more effectively to this coefficient at moderate and high temperatures, further amplifying the growth of $D_{F}/T^{2}$. Physically, this indicates that the chemical potential gradients can induce substantial heat transport when the system is baryon-rich, reflecting the strong coupling between baryon and heat currents in such media. Furthermore, in the hadronic medium, a strong dependence of temperature can be observed for both the Diffusion and Dufour coefficients. For $\mu_{B}$ = 0.20 GeV, we observe an increasing trend which saturates around $T \simeq 0.13 $ GeV. Meanwhile, for a slightly higher chemical potential of 0.40 GeV, we get competing effects from $f_i^0 (1 \pm f_i^0)$ and $\tau_R^i$, thus resulting in a slight dip at $T \simeq 0.15 $ GeV. Moreover, the $\tau_R^i$ dominates for $\mu_{B}$ = 0.60 GeV, and as a result, we observe a decreasing trend. However, in the QGP phase, the temperature dependence of $D/T$ and $D_{F}/T^{2}$ is weak due to the fact that $\tau_R^i \sim 1/T$, which can be seen from Eq.~\ref{tauqgp}. Additionally, one can notice that the enhancing effect of baryon chemical potential on $D/T$ and $D_{F}/T^{2}$ is still prominent in the QGP medium.

On the other hand, the Soret coefficient, expressed in Eq.~(\ref{Eq:Soret}), does not contain an explicit $b_{i}$ weighting factor. Although baryonic species contribute through the $(\omega_{i}-b_{i}\mu_{B})$ term, the overall magnitude and temperature dependence of $S_{T}/T$ are largely governed by the light mesonic species ($b_{i}=0$), such as pions, which remain abundant even at very small $\mu_{B}$. With an increase in $\mu_{B}$, we observe a decrease in $S_{T}/T$ at low temperature. This is due to the presence of the factor of $(\omega_{i}-b_{i}\mu_{B})~\tau_R^i$, which decreases with increasing $\mu_{B}$. As the system approaches the crossover region and transitions into the QGP phase, the Soret coefficient shows an increase with temperature due to the liberation of quark degrees of freedom, which carry baryon number but have much smaller effective masses. Furthermore, compared to the Dufour coefficient, the sensitivity of the Soret coefficient to $\mu_{B}$ remains relatively weak, indicating that temperature gradients, rather than baryon density, primarily drive its behavior in both phases. In contrast to the Diffusion and Dufour coefficients of the hadronic medium, we observe a large Soret coefficient at a lower baryon chemical potential. This hints towards the fact that the diffusion of baryons, due to the Soret effect, will be most affected by the temperature gradient in the nearly baryon-free region.
A similar trend is observed for the thermal conductivity, $\kappa/T^{2}$, as described by Eq.~(\ref{Eq:kappa}), due to the presence of the quadratic energy-weighting factor $(\omega_{i}-b_{i}\mu_{B})^{2}$. We observe a similar temperature and baryon chemical potential dependence of $\kappa/T^{2}$ across the entire temperature range. 
\section{Summary}
In summary, for the very first time, we investigate the temperature and baryon chemical potential dependence of coupled-transport coefficients of strongly interacting matter, namely the Soret coefficient and the Dufour coefficient, along with the baryon diffusion coefficient and the thermal conductivity, within the framework of a hadron resonance gas model for the hadronic phase and a quasiparticle description of the quark-gluon plasma phase. Our analysis reveals that the diffusion and Dufour coefficients exhibit a strong sensitivity to the baryon chemical potential due to their explicit linear weighting with the baryon number, leading to a significant enhancement of baryon-driven transport in baryon-rich environments. The baryon-chemical potential dependence of the Dufour coefficient highlights its importance in low-energy heavy-ion experiments, such as those at the Nuclotron-based Ion Collider facility (NICA). By affecting the heat current and, consequently, the cooling rate, the Dufour effect could offer deeper insights into the thermodynamic evolution of baryon-rich QCD matter.

In a strongly interacting QCD medium, heat conduction mainly arises through two key mechanisms. The first is the frequent interactions among the constituents of the medium, which facilitate the transfer of energy. The second is the role of thermal conductivity, which enhances the random motion of these particles and accelerates the redistribution of energy across the system. The coupled transport also plays an important role in linking baryon and heat transport in such a medium. These effects are crucial in several fields, including nuclear reactor design, geothermal energy systems, groundwater pollution studies, oil reservoir management, isotope separation, manufacturing of rubber and plastic sheets, gas mixing, compact heat exchangers, and the safe disposal of nuclear waste~\cite{shaheen2021soret, Gajjela:2020abc}. The wide-ranging applications highlight the importance of understanding and quantifying the interplay between thermal and mass transport driven by the Soret and Dufour effects in both natural and engineered systems. In the context of heavy-ion collisions, a detailed understanding of the Soret and Dufour effects is essential for capturing the coupled transport of baryon number and heat in the evolving QCD medium. These coupled-transport mechanisms can significantly influence the redistribution of baryon density and thermal gradients during the fireball's evolution, maybe affect the key observables such as net-baryon fluctuations, particle spectra, and flow anisotropies across varying beam energies. Our findings of Soret and Dufour effects in QCD matter open new avenues for understanding coupled heat and baryon transport in strongly interacting systems. These insights are not only crucial for interpreting heavy-ion collision dynamics but also for modeling astrophysical environments such as neutron stars or magnetars, where extreme conditions prevail. Incorporating the influence of strong magnetic fields in future work could uncover anisotropic transport phenomena, further enriching the theoretical and phenomenological landscape of QCD matter.

\acknowledgments
K.S. acknowledges the doctoral fellowship from the UGC, Government of India. K.G. acknowledges the financial support from the Prime Minister's Research Fellowship (PMRF), Government of India. R.S. gratefully acknowledges the DAE-DST, Govt. of India funding under the mega-science project – “Indian participation in the ALICE experiment at CERN” bearing Project No. SR/MF/PS-02/2021-IITI (E-37123). 

\end{document}